\documentclass[aps,pre,twocolumn,showpacs,groupedaddress]{revtex4}
\usepackage{graphicx}
\begin{document}

\title{Effect of Equilibration on Primitive Path Analyses of Entangled 
Polymers}
\author{Robert S. Hoy}
\email{robhoy@pha.jhu.edu}
\author{Mark O. Robbins}
\affiliation{Department of Physics and Astronomy, Johns Hopkins University, 
Baltimore, Maryland 21218}
\date{October 22, 2005}
\begin{abstract}
We use recently developed primitive path analysis (PPA) methods to study the 
effect of equilibration on entanglement density in model polymeric systems.  
Values of $N_{e}$ for two commonly used equilibration methods differ by a 
factor of two to four even though the methods produce similar large-scale 
chain statistics.
We find that local chain stretching in poorly equilibrated samples increases 
entanglement density.
The evolution of $N_{e}$ with time shows that many entanglements are lost 
through fast processes such as chain retraction as the local stretching 
relaxes.
Quenching a melt state into a glass has little effect on $N_{e}$.
Equilibration-dependent differences in short-scale structure affect the craze 
extension ratio much less than expected from the differences in PPA values of 
$N_{e}$.
\end{abstract}
\maketitle

\section{Introduction}
\label{sec:Intro}
Much of the physics of bulk polymeric material is independent of chemical 
detail and arises instead from the connectivity of the constituent chain 
molecules, provided the material is sufficiently dense and the chains 
sufficiently long.
Since chains can not pass through each other, they are confined to ``tubes'' 
by topological constraints that are often associated with discrete 
``entanglements''.
The tube model of Edwards and de Gennes has been a focus of extensive 
study over the past few decades \cite{doi88, mcleish02} and 
qualitatively describes many aspects of entangled polymer dynamics.  
Perhaps its most famous prediction is the relation $\eta \propto N^{3}$ for 
$N \gg N_{e}$, where $\eta$ is the melt viscosity, $N$ is the degree of 
polymerization, and $N_{e}$ is the ``entanglement length''.

A key construct of the tube model is the ``primitive path", the shortest 
contour into which a chain molecule fixed at its ends may contract without 
crossing any other chains \cite{doi88}.
Everaers and collaborators have recently used this construct to perform 
``primitive path analysis'' (PPA) of model entangled polymeric systems 
\cite{everaers04, sukumaran05}.
Their results for $N_{e}$ agree with the predictions of a chain packing 
model \cite{fetters94} that explains trends in experimental results for melts 
and semidilute solutions of linear homopolymers 
\cite{fetters99b,fetters99,fetters94,colby91,inoue02}.
The PPA technique is a promising tool for obtaining information about the 
behavior of individual entanglements which has not been accessible through 
other theoretical or experimental means.

In this paper, we address the relation of melt equilibration to the 
entanglement length measured by  PPA analysis.  
Experimental studies of linear homopolymer melts have shown that $N_{e}$ is 
controlled by the Gaussian statistics of chains at large scales 
\cite{fetters99b,fetters99,fetters94}.
One might reasonably assume that values of $N_{e}$ from PPA analysis of 
model melts would also be controlled by large-scale chain structure.
Our results, however, show that this assumption can fail badly.
Well-equilibrated states produced using a double-bridging algorithm 
\cite{auhl03} are compared to states produced using the common 
fast-pushoff method.
For typical equilibration times, use of the fast pushoff produces chain 
stretching at short scales while preserving large-scale structure 
\cite{auhl03}.
This stretching can produce values of $N_e$ that are two to four times 
smaller than those for well-equilibrated states, showing that failure to 
accurately equilibrate  \textit{short-scale} chain structure can cause large 
systematic errors in PPA analyses.

Our values of $N_{e}$ for the well-equilibrated states agree with values 
from previous PPA analyses \cite{everaers04,sukumaran05,zhou05} and from other 
methods \cite{puetz00}.
We verify that the reduction of $N_{e}$ for poorly equilibrated states 
reflects a real excess of topological constraints, and show that early stages 
of the evolution of $N_{e}$ towards equilibrium can occur through fast 
topology-changing processes familiar from tube theories \cite{mcleish02},
such as cooperative chain retraction and constraint release.
The changes in $N_{e}$ with improving equilibration are correlated to 
changes in chain structure.

We also perform PPA analyses of glassy states.
In interpreting the mechanical properties of polymer glasses, it is often 
assumed that the value of $N_{e}$ in the glass is inherited from the melt 
\cite{haward97,kramer90}.
While PPA values for $N_{e}$ drop slightly upon cooling from a melt to 
a glass well below $T_{g}$, the changes are comparable to systematic and 
statistical uncertainties.  
Glass values for $N_e$ can not be compared to rheological measurements, 
but can be obtained from the extension ratio during craze formation 
\cite{kramer90,rottler03}.
We find that this measure of $N_e$ is less sensitive to equilibration than
PPA values, perhaps because deformation into the craze structure removes 
excess entanglements produced by local stretching.

\section{Polymer Model: Methods of Melt Equilibration}
\label{sec:modelmethod}
We employ a coarse-grained bead-spring polymer model 
\cite{bishop80, kremer90} that incorporates key physical features of linear 
homopolymers such as covalent backbone bonds, excluded-volume 
interactions, chain stiffness, and the topological restriction that chains 
may not cross.
All monomers have mass $m$ and interact via the truncated and shifted
Lennard-Jones potential:
\begin{equation}
\begin{array}{lll}
U_{LJ}(r) & = & 4\epsilon\left(\left((\frac{a}{r})^{12} - 
(\frac{a}{r_{c}})^{12}\right) - \left((\frac{a}{r})^{6} - 
(\frac{a}{r_{c}})^{6}\right)\right),
\end{array}
\label{eq:shiftedLennardJones}
\end{equation}
where $r_{c}$ is the potential cutoff radius and $U_{LJ}(r) = 0$ for 
$r > r_{c}$.  Unless noted, $r_{c} = 2^{1/6}a$.  
We express all quantities in terms of length $a$, energy $\epsilon$, 
and time $\tau_{LJ} = \sqrt{ma^{2}/\epsilon}$. 

Covalent bonds between adjacent monomers on a chain are modeled using 
the finitely extensible nonlinear elastic (FENE) potential 
\begin{equation}
\begin{array}{lll}
U_{FENE}(r) & = & -\displaystyle\frac{kR_{0}^2}{2}\rm ln(1 - (r/R_{0})^{2}),
\end{array}
\label{eq:FENEbondpotential}
\end{equation}
with the canonical parameter choices $R_{0} = 1.5a$ and 
$k = 30\epsilon/a^{2}$ \cite{kremer90}.  The chains contain $N$ monomers and 
have an equilibrium bond length $l_{0} = .96a$.  
Chain stiffness is modeled using the potential
\begin{equation}
\begin{array}{lll}
U_{bend}(r) & = & k_{bend}\left(1 - \displaystyle\frac{\vec{b}_{i-1}
\cdot\vec{b}_{i}}{|\vec{b}_{i-1}||\vec{b}_{i}|}\right),\\
 & = & k_{bend}(1 - cos(\theta_{i})),
\end{array}
\label{eq:ChainStiffnessPotential}
\end{equation}
where $\vec{r}_{i}$ denotes the position of the ith monomer on a chain and 
$\theta_{i}$ is the angle betweeen consecutive bond vectors 
$\vec{b}_{i-1} = \vec{r}_{i} - \vec{r}_{i-1}$ and 
$\vec{b}_{i} = \vec{r}_{i+1} - \vec{r}_{i}$.   

Newton's equations of motion are integrated with the velocity-Verlet method 
\cite{frenkel02} and timestep $\delta t = .005\tau_{LJ}-.012\tau_{LJ}$.  
The system is coupled to a heat bath at temperature $T$ using a Langevin 
thermostat \cite{schneider78}.  
Unless noted, the simulation cell is a cube whose size is chosen so that the 
monomer number density is $\rho = 0.85a^{-3}$.
Periodic boundary conditions are applied in all three directions.
At this density and $T = 1.0\epsilon/k_{B}$ the system is a melt well above the
glass transition \cite{kremer90}.
For this model, published entanglement lengths vary from about 70 for fully
flexible chains $(k_{bend} = 0)$ to 20 for semiflexible chains with 
$k_{bend} = 2.0\epsilon$
\cite{everaers04,sukumaran05,zhou05,faller01,puetz00,rottler03}.  
For the chain lengths used in this study ($N = 350$ and $N=500$), larger 
values of $k_{bend}$ produce nematic order \cite{faller99}.

We contrast two common methods for creating initial states for the PPA 
analysis.  
Both begin by generating initial configurations of $N_{ch}$ chains without 
considering excluded volume \cite{theodorou85}.
Each initial chain configuration is a random walk of $N-1$ steps with the bond 
angles chosen to give the desired chain statistics \cite{auhl03} 
\begin{equation}
\begin{array}{l}
<R^{2}(n)> = nl_{0}^{2}\bigg(\displaystyle\frac{1 + <cos(\theta)>}{1 - <cos(\theta)>}\\ -
\displaystyle\frac{2}{n}\displaystyle\frac{<cos(\theta)>
(1 - <cos(\theta)>^{n})}{(1-<cos(\theta)>)^{2}}\bigg),
\end{array}
\label{eq:Gaussianwithcorrection}
\end{equation}
where $<R^{2}(n)>$ is the average squared distance between monomers separated
 by chemical distance n.  
Large scale measures of chain structure such as the chain stiffness constant
 $C_{\infty}$ and Kuhn length $l_{K}$ are related to the chain statistics by
\begin{equation}
C_{\infty} = \displaystyle\frac{l_{K}}{l_{0}} = \displaystyle
\frac{1 + <cos(\theta)>}{1 - <cos(\theta)>}.
\label{eq:chainstiffnessconst}
\end{equation}

Lennard-Jones interactions for nonbonded monomers cannot be introduced 
immediately after creating the initial chain configurations because chains
 spatially overlap \cite{theodorou85}.    
Instead, a soft repulsive potential is used to reduce the chain overlap 
through gradual introduction of excluded volume interactions.  
The specific soft potential used is 
\begin{equation}
U_{soft}(r) = \bigg\{ \begin{array}{ccc}
A\left[1 + cos\left(\frac{\pi r}{2^{1/6}a}\right)\right] & , & r < 2^{1/6}a\\
0 & , & r > 2^{1/6}a
\end{array}
\label{eq:softpushoff}
\end{equation}
for nonbonded beads.  The value of A is linearly increased from $4\epsilon$ 
to $100\epsilon$ over a time $20\tau_{LJ}$, as in Ref.\ \cite{auhl03}.  
Unfortunately, this ``fast pushoff'' procedure creates significant distortions 
in the chain statistics on length scales comparable to the tube diameter 
\cite{auhl03,brown94}.  Chains are stretched well beyond the form of 
Eqn.\ (\ref{eq:Gaussianwithcorrection}) at intermediate chemical distances 
\cite{auhl03}.

After the fast pushoff is completed, the two methods of equilibration differ.  
In the pure molecular dynamics (PMD) method, normal Lennard-Jones
interactions are activated, and integration is continued for up to 
$4.5\cdot10^{5}\tau_{LJ}$. 
This maximum equilibration time is still much shorter 
than the longest relaxation time of the systems we consider, which is the 
disentanglement time $\tau_{d}$.  
Only after an equilibration run of more than $\tau_{d}$ do the chain 
configurations come to full equilibrium \cite{auhl03}, but equilibration 
runs much shorter than $\tau_{d}$ are commonly used in PMD simulations 
\cite{kremer90, auhl03}.  
We therefore refer to our PMD-prepared states as ``poorly equilibrated''.

The other equilibration method used is the double-bridging-MD hybrid (DBH) 
algorithm described in Ref.\ \cite{auhl03}. 
In addition to standard MD equilibration, Monte Carlo moves which alter 
the connectivity of chain subsections are periodically performed, allowing 
the chain configurations to relax far more rapidly \cite{karayiannis02}.
Equilibrated chain statistics from extremely long MD runs were used to obtain 
target functions for these Monte Carlo moves \cite{auhl03,puetz00}.
We therefore refer to the DBH-prepared states as ``well equilibrated''.

If a glassy state is desired, we increase $r_{c}$ to $1.5a$ and perform a 
rapid temperature quench at a cooling rate of  
$\dot{T} = -2\times10^{-3}(\epsilon/k_{B})/\tau_{LJ}$.
The systems are cooled to $T = .55\epsilon/k_{B}$ at constant density, and then
 to $T = .1\epsilon/k_{B}$ at zero pressure using a Nose-Hoover barostat 
\cite{frenkel02}.  The resulting glasses have density $\rho \simeq 1.02a^{-3}$.
Other temperature and pressure protocols give similar results for $N_{e}$.

Once preparation of the system is complete, we perform primitive path analyses
as in Refs.\ \cite{everaers04,sukumaran05}.
Except as noted in Section \ref{sec:results}, the procedure is nearly 
identical to that of Ref.\ \cite{sukumaran05}, and we refer to it as the 
``standard'' PPA procedure.  All chain ends are fixed in space and several 
changes are made to the interaction potential.
Intrachain excluded-volume interactions are deactivated, while interchain 
excluded-volume interactions are retained.
The covalent bonds are strengthened by setting $k = 100\epsilon$, and the bond
 lengths are capped at $1.2a$ to prevent chains from crossing one another 
\cite{sukumaran05}.
Note that we do not attempt to preserve self-entanglements, since their 
number is negligibly small for the systems considered here \cite{sukumaran05}.
For semiflexible chains, the bond-bending potential is deactivated by
 setting $k_{bend} = 0$.
The system is then coupled to a heat bath at $T = .001\epsilon/k_{B}$ so that
thermal fluctuations are negligible, and the equations of motion are integrated
 until the chains minimize their length.  
This typically requires from 500 to 2000 $\tau_{LJ}$.
Other variants of the PPA procedure are discussed in Section 
\ref{subsec:notfric}.

Once the chain contour lengths have been minimized, we use the formula given 
in Ref.\ \cite{everaers04} to calculate the entanglement length:
\begin{equation}
N_{e} = \displaystyle\frac{<R^{2}_{ee}>}{(N-1)<b_{pp}>^{2}},
\label{eq:standardNe}
\end{equation}
where $<R^{2}_{ee}>$ is the average squared end-end distance, and 
$<b_{pp}>$ is the mean bond length at the end of the PPA run
 \cite{everaers04}.
We also calculate the rms variation of $b_{pp}$ for each chain, and report the
mean of this quantity as $\Delta b_{pp}$.  
An alternative method for calculating $N_{e}$ is to fit the primitive path 
chain statistics to Eqn.\ (\ref{eq:Gaussianwithcorrection}), with $l_{0}$ 
and $<cos(\theta)>$ as fitting parameters \cite{sukumaran05}.  
The fit value of $<cos(\theta)>$ is then inserted into Eqn.\ 
\ref{eq:chainstiffnessconst}, and $N_{e}$ is identified with $C_{\infty}$
 \cite{sukumaran05}.
In every case this gave values of $N_{e}$ consistent with the values from 
Eqn.\  (\ref{eq:standardNe}), confirming that the primitive paths have 
Gaussian random walk statistics, with $N_{e}$ monomers per Kuhn segment 
\cite{everaers04}.

\section{Results}
\label{sec:results}
\subsection{Dependence of $N_{e}$ on Preparation Method}
\label{subsec:notfric}
Table \ref{tab:NeTabflexsemiflex} shows results from PPA runs for flexible 
$(k_{bend} = 0)$ and semiflexible ($k_{bend} = 0.75\epsilon$ or $1.5\epsilon$)
melt states prepared with the PMD and DBH methods.
These results are for $N_{ch} = 500$ chains of length $N = 500$, so 
finite-size effects are small \cite{sukumaran05}.
The measured entanglement lengths depend dramatically on equilibration 
procedure.
For states prepared using the DBH method, our results for $N_{e}$ agree with 
values from Ref.\ \cite{everaers04}.
$N_{e}$ is smaller by a factor of two to three for the poorly-equilibrated, 
PMD-prepared states.  
 
\begin{table}[h]
\caption{Primitive path analysis results for flexible and semiflexible melts -
standard procedure.  All of the PMD states listed were equilibrated for
$480\tau_{LJ}$ following the fast pushoff.  The quoted  uncertainties in 
$N_{e}$ are the errors on the means of the distributions of 
$R^{2}_{ee}/((N-1)b_{pp}^2)$ for the $N_{ch}$ chains, with each chain 
considered an independent measurement.}
\begin{ruledtabular}
\begin{tabular}{lcccc}
Method & $k_{bend}/\epsilon$ & $N_{e}$ & $<b_{pp}>/a$ & $\Delta b_{pp}/a$\\
DBH & 0 &$73.2\pm2.4$ & 0.1543 & 0.0037\\
PMD & 0 & $24.9\pm0.9$ & 0.2695 & 0.0144\\
DBH & 0.75 & $45.1\pm1.5$ & 0.2135 & 0.0042\\
PMD & 0.75 & $18.3\pm0.6$ & 0.3397 & 0.0304\\
DBH & 1.5 & $28.1\pm1.0$ & 0.3078 & 0.0119\\
PMD & 1.5 & $13.1\pm0.5$ & 0.4510 & 0.0612
\end{tabular}
\end{ruledtabular}
\label{tab:NeTabflexsemiflex}
\end{table}

Another difference we find is that $\Delta b_{pp}$ is larger for poorly 
equilibrated initial states.  Fluctuations in $b_{pp}$ indicate that friction 
between chains has prevented stress equilibration along the chains.  
This alone would not change the value of $N_{e}$ from Eqn.\ 
(\ref{eq:standardNe}).  
However, $N_{e}$ would be decreased if friction prevented the chains from 
minimizing the total contour length, for example by trapping free loops along 
chains.

To test the potential magnitude of such effects we explored different 
algorithms for obtaining the primitive paths.  
These included beginning the PPA procedure at $T=1.0\epsilon/k_{B}$ and cooling slowly, 
reducing the excluded volume interaction between adjacent monomers gradually, 
and increasing the value of $a$ in the LJ potential to reduce friction between 
chains.  
Ref.\ \cite{sukumaran05} specifies $\Delta b_{pp} < .006a$ as a criterion 
for convergence of PPA runs to a state with uniform bond tension in the 
individual chains. 
For the standard PPA procedure, this convergence failed to occur for four of 
the six systems in Table \ref{tab:NeTabflexsemiflex}.
Table \ref{tab:NeTabflexsemiflexB} shows results for these states with our 
altered PPA procedure.
While improved stress equilibration reduces $\Delta b_{pp}$ by a factor of two,
 the values of $N_{e}$ increase by at most five percent.  
Thus it seems interchain friction is not responsible for the large 
differences in $N_{e}$.

\begin{table}[h]
\caption{Primitive path analysis results for flexible and semiflexible melts 
with procedure modified to reduce friction.  Initial 
states and uncertainties are as in Table \ref{tab:NeTabflexsemiflex}.}
\begin{ruledtabular}
\begin{tabular}{lcccc}
Method & $k_{bend}/\epsilon$ & $N_{e}$ & $<b_{pp}>/a$ & $\Delta b_{pp}/a$\\
PMD & 0 & $26.1\pm0.9$ & 0.2638 & 0.0072\\
PMD & 0.75 & $18.9\pm0.6$ & 0.3346 & 0.0143\\
DBH & 1.5 & $28.8\pm1.0$ & 0.3040 & 0.0067\\
PMD & 1.5 & $13.3\pm0.5$ & 0.4478 & 0.0286 
\end{tabular}
\end{ruledtabular}
\label{tab:NeTabflexsemiflexB}
\end{table}

The entanglement length of a bulk polymeric system is often related to 
large-scale measures of chain structure such as the packing length p 
\cite{fetters94}:
\begin{equation}
p = \displaystyle\frac{N}{\rho<R_{ee}^{2}>},
\label{eq:packinglength}
\end{equation}
which is the volume occupied by a chain divided by its mean-square end-end 
distance.
Extensive experiments on linear, Gaussian-chain homopolymers have shown 
that $N_{e} \propto \rho p^{3}$ \cite{fetters94,fetters99b,fetters99}.   
The packing lengths are $.68a$ and $.65a$ for the DBH-prepared and 
PMD-prepared $k_{bend} = 0$ states, respectively.  
Based on the experimentally observed power-law dependence, one might 
expect that the $N_{e}$ for the two systems should differ by only about 
10-15\%.
The actual difference in the $N_{e}$ measured by PPA analyses is a factor of 
three, as noted above.
Similarly, the differences in $N_{e}$ between well- and poorly-equilibrated 
states for $k_{bend} = 0.75\epsilon$ and $k_{bend} = 1.5\epsilon$ are far 
too large to be explained by any difference in their large-scale structure.

\begin{figure}[htbp]
\includegraphics[width=3.375in]{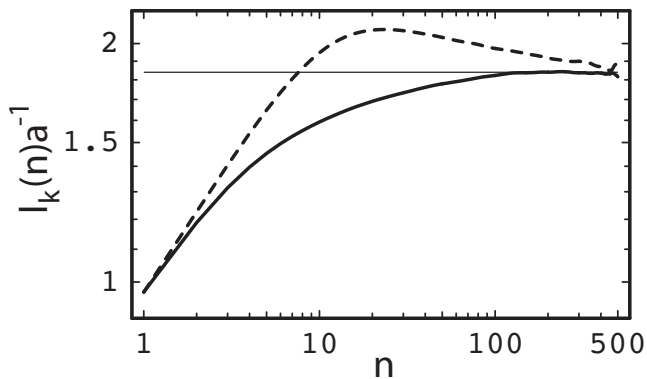}
\caption{Chain statistics $l_{k}(n)$ for PMD (dashed curve) and DBH (solid 
curve) melt states of 500 flexible chains of length N = 500.  The 
horizontal line, $l_{k}(n) = 1.84a$, illustrates the large-n Gaussian limit 
of the well-equilibrated state.} 
\label{fig:stretchinginflex500500}
\end{figure}

The differences in $N_{e}$ may, however, be partially understood in terms of 
differences in \textit{short-scale} structure.  
Figure \ref{fig:stretchinginflex500500} contrasts the chain statistics for 
the PMD and DBH flexible melt states, prior to primitive path analysis.  
To characterize chain configurations, we find it useful to define the 
Kuhn-length-like quantity 
\begin{equation}
l_{k}(n) \equiv \frac{<R^{2}(n)>}{nl_{0}}.
\label{eq:locallk}
\end{equation}
This quantity was found to be more sensitive to equilibration than the other
quantities considered in Ref.\ \cite{auhl03}.
In Figure \ref{fig:stretchinginflex500500}, $l_{k}(n)$ for the DBH-prepared 
state increases nearly monotonically with n, displaying a Gaussian limit 
$l_{k}(n) \simeq l_{K} = 1.84a$ for chemical distances $n > 100$.  
The PMD states show pronounced non-Gaussian behavior. 
Results for $l_{k}(n)$ rise too steeply at small n, reach a peak at 
$n = n_{max}$, then drop to essentially the same large scale value as the DBH 
state as $n\to N$ \cite{auhl03}.
The differences in the short-scale chain configurations are qualitatively 
consistent with the difference in $N_{e}$; the straightening of chains at 
small scales in the PMD state is similar to that produced by increasing 
$k_{bend}$, which also lowers $N_{e}$ \cite{haward97,fetters94,faller01}.

\begin{figure}[htbp]
\includegraphics[width=3.375in]{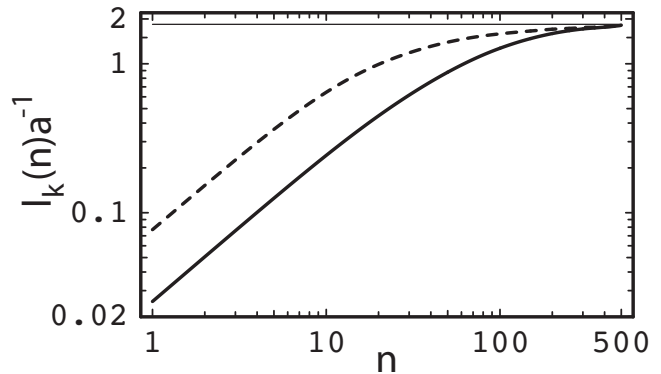}
\caption{Chain statistics $l_{k}(n)$ for PMD (dashed curve) and DBH (solid 
curve) melt states of 500 flexible chains of length N = 500, after 
implementation of the PPA.} 
\label{fig:flex500500postPPA}
\end{figure}

Figure \ref{fig:flex500500postPPA} contrasts the chain statistics of the 
primitive paths for the PMD and DBH flexible melt states.  
The peak in the PMD statistics at intermediate range is suppressed and the 
final curve is consistent with Gaussian statistics.
This is why we get the same $N_{e}$ from Eqn.\ (\ref{eq:standardNe}) or a fit 
to the full curve.
The change in the PMD statistics during implementation of the PPA appears to 
be more dramatic than that for the DBH state.  
However, the relative displacement of monomers at $n > N_{e}$ is only of 
order $a$, i.e.\ small compared to the tube diameter, in both cases.

\subsection{Evolution of $N_{e}$ with PMD Equilibration Time}
\begin{figure}[htbp]
\begin{center}
\includegraphics[width=3.375in]{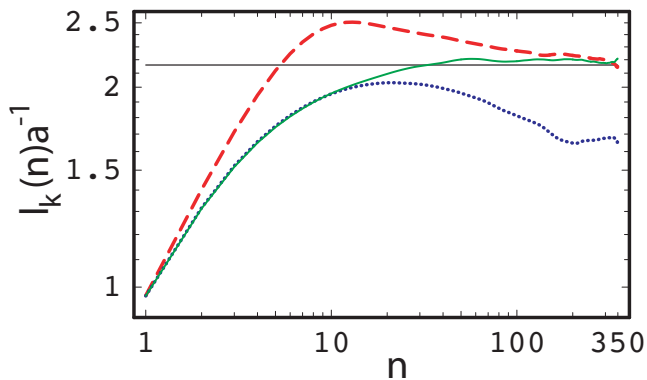}
\end{center}
\caption{Evolution of melt chain statistics $l_{k}(n)$ with 
PMD equilibration time for a 200-chain N = 350 system with 
$k_{bend} = 0.75\epsilon$ at $T = 1.0\epsilon/k_{B}$.  
The dashed and dotted curves are after $500\tau_{LJ}$ and 
$2\cdot10^5\tau_{LJ}$ (about one Rouse time), respectively.  
The solid curve shows chain statistics of a well-equilibrated DBH state, and 
the thin horizontal line corresponds to the equilibrium Kuhn length, 
$l_{K} = 2.16a$.}
\label{fig:chainstats-timedependence}
\end{figure}

The effect of short-scale structure on $N_{e}$ can be further examined by
following both during sample equilibration. 
Figure \ref{fig:chainstats-timedependence} shows how the melt chain 
statistics evolve with increasing PMD equilibration time for a semiflexible
($k_{bend} = 0.75\epsilon$) $N_{ch} = 200$, $N = 350$ system.
After equilibration runs of order $10^{3}\tau_{LJ}$ following the fast 
pushoff, the chain statistics show the short scale stretching typical of 
poorly equilibrated states \cite{auhl03}.  
After an equilibration run of one Rouse time, 
$\tau_{R} \sim 2\cdot10^{5}\tau_{LJ}$ \cite{puetz00}, the chain statistics 
are well equilibrated at chemical distances up to about ten monomers, 
but still out of equilibrium at chemical distances approaching the chain 
length. 
Table \ref{tab:equilibtimedep} shows additional information on the pre-PPA 
melt chain statistics, as well as PPA results for this system for PMD 
equilibration times up to $4.5\cdot10^{5}\tau_{LJ} \sim 2\tau_{R}$.
Results for well-equilibrated DBH-prepared states from Ref.\ \cite{everaers04} 
are shown for comparison.
The value of $N_e$ rises rapidly, doubling by $\tau_{R}/4$ and tripling
by $\sim \tau_{R}$.
Note that the values of $\Delta b_{pp}$ drop with increasing $t_{eq}$ over 
roughly the same timescale, indicating that friction effects diminish. 

\begin{table*}[htbp]
\caption{Evolution of PPA results for a $N_{ch} = 200$, $N = 350$ semiflexible 
($k_{bend} = 0.75\epsilon$) system with PMD melt equilibration time $t_{eq}$, 
following a fast pushoff.
The Rouse time of these chains is $\tau_{R} \sim 2\cdot 10^{5}\tau_{LJ}$ 
\cite{puetz00}.  
Error bars are as defined in Table \ref{tab:NeTabflexsemiflex}.
In the bottom row, $n_{max}$, $l_{k}$, and $<R_{ee}^{2}>$ are limiting values 
for 
$N_{ch} \to \infty$, while $N_{e}$, $b_{pp}$, and $\Delta b_{pp}$ are from 
Ref.\ \cite{everaers04}.}
\begin{ruledtabular}
\begin{tabular}{lcccccc}
$t_{eq}/\tau_{LJ}$ & $n_{max}$ & $l_{k}(n_{max})/a$ & 
$10^{-2}<R_{ee}^{2}>/a^{2}$ & $N_{e}$ & $<b_{pp}>/a$ & $\Delta b_{pp}/a$\\
$5\cdot10^{2}$ & 13 & 2.51 & 7.20 & $11\pm1$ & 0.434 & 0.058\\
$3\cdot10^{3}$ & 18 & 2.39 & 6.89 & $12\pm2$ & 0.405 & 0.048\\
$5\cdot10^{4}$ & 19 & 2.24 & 6.79 & $22\pm2$ & 0.300 & 0.016\\
$1.0\cdot10^{5}$ & 21 & 2.16 & 6.22 & $25\pm2$ & 0.265& 0.010\\
$2.0\cdot10^{5}$ & 21 & 1.99 & 5.55 & $31\pm2$ & 0.228 & 0.005\\
$3.0\cdot10^{5}$ & 26 & 2.04 & 5.82 & $34\pm3$ & 0.220 & 0.004\\
$4.5\cdot10^{5}$ & 29 & 2.00 & 6.48 & $43\pm3$ & 0.207 & 0.003\\
Equilibrated & 350 & 2.16 & 7.24  & $45\pm3$ & 0.21 & $<0.006$
\end{tabular}
\end{ruledtabular}
\label{tab:equilibtimedep}
\end{table*}

The changes in $N_{e}$ and $l_{k}$ described above are much faster
than the disentanglement time; $\tau_{R}$ is only about $\tau_{d}/5$ for
this system \cite{puetz00}.
Thus the loss 
of entanglements must occur through a mechanism which is faster than
reptation.
Chain retraction is a mechanism studied in nonlinear-response tube theories
 \cite{mcleish02} wherein the unentangled ends of a stretched chain contract 
rapidly inwards along its tube, thereby shortening the tube and releasing
entanglement constraints.
This process is much faster than reptation for $N/N_{e} \gg 1$ because motion 
of the chain ends arises from entropic tension rather than stochastic 
diffusion, and because the chain center of mass need not move 
\cite{mcleish02}.
Since the fast pushoff stretches chains, chain retraction begins as soon as the 
Lennard-Jones interactions are activated.
As shown in Figure \ref{fig:chainstats-timedependence} and Table 
\ref{tab:equilibtimedep}, the chains rapidly contract at scales comparable to 
the end-end distance.
$<R_{ee}^{2}>$ drops as much as four standard deviations below its equilibrium 
value at $t_{eq} \simeq \tau_{R}$, showing that this contraction is not an 
equilibrium fluctuation.

As most of the chains are contracting in this way simultaneously, 
many topological constraints are also released far from chain ends.
The combination of chain retraction and constraint release \cite{mcleish02} 
accounts for the observed evolution of the chain statistics at small 
chemical distances.
Another indication of the presence of constraint release is that the 
relaxation of the primitive path length $L_{pp} = (N-1)<b_{pp}>$, which is 
well fit by the tube-theory prediction for chain retraction \cite{doi88}
\begin{equation}
L_{pp}(t_{eq}) = L_{0} + \displaystyle\frac{L_{1}}{\sqrt{t_{eq}}} + 
L_{2}\sum_{\text{odd p}}\displaystyle\frac{\exp(-p^{2}t_{eq}/\tau_{pp})}{p^{2}},
\label{eq:primpathlengthdecay}
\end{equation}
has a decay time which is clearly below 
$\tau_{R}$; $\tau_{pp}\simeq8\cdot10^{4}\tau_{LJ} \sim 0.4\tau_{R}$.
Studies with different initial states and $k_{bend}=0$ also show that 
entanglement loss is more rapid than reptation, with $\tau_{pp}$ about half 
of $\tau_{R}$.

As noted in Section \ref{sec:modelmethod}, full equilibration of $N_{e}$ 
should occur only after the disentanglement time $\tau_{d}$, when the 
chains have vacated their original tubes.
It is interesting that the $N_{e} = 43$ value obtained for the 
longest-equilibrated PMD state is consistent with values obtained for 
well-equilibrated states \cite{everaers04}, when clearly the chain 
configurations remain far from equilibrium at large chemical distances.
This suggests that short-range order plays the most important role
in determining entanglements.
Confirming this would require following the evolution of $N_{e}$
to $t > \tau_{d}$, which is beyond the scope of this study.

\subsection{Primitive Path Analyses of Glasses}
\label{subsec:glasPPAs}
The six melt states from Table \ref{tab:NeTabflexsemiflex} were rapidly 
quenched, and standard primitive path analyses were applied to the 
resulting glassy states.
In every case, the value of $N_{e}$ in the glass was close to that in the
corresponding melt, even though melt values depended strongly on $k_{bend}$
and equilibration.
It is often assumed \cite{kramer90,haward97} that glasses inherit the melt 
value for $N_{e}$, and our results are consistent with this assumption to 
within statistical and systematic errors ($\sim$5\%).

Nevertheless, it is interesting that the values of $N_{e}$ were lower in the 
glass than in the melt in all six cases.  
This reduction could in principle arise from an increase in interchain friction 
associated with the higher density of the glassy states, but the values of 
$\Delta b_{pp}$ in Table \ref{tab:NeTabglasses} rule this out.  
All were remarkably close to the values for the corresponding 
melt states in Table \ref{tab:NeTabflexsemiflex}, indicating that friction is 
not responsible for the changes in $N_{e}$.

\begin{table*}[htbp]
\caption{Primitive path analysis results for glasses.    
The quoted uncertainties in $N_{e}$ (PPA) are as in Table 
\ref{tab:NeTabflexsemiflex}.  
Values of $\lambda^{th}$ use the PPA values of $N_{e}$, with 
$\lambda^{th} = \sqrt{N_{e}l_{0}/l_{K}}$.  Values of $\lambda$ are measured 
using the same procedure as Ref.\ \cite{rottler03}, and $N_{e}^{th} = 
\lambda^{2}l_{K}/l_{0}$.} \begin{ruledtabular}
\begin{tabular}{lcccccc}
Method & $k_{bend}/\epsilon$ & $N_{e}$ (PPA) & $\Delta b_{pp}/a$ & 
$\lambda^{th}$ & $\lambda$ & $N_{e}^{th}$\\
DBH & 0 &$71.2\pm2.3$ & 0.0038 & $6.5\pm0.2$ & $5.9\pm0.4$ & $59\pm8$\\
PMD & 0 & $23.3\pm0.9$ & 0.0126 & $3.6\pm0.2$ & $5.9\pm0.2$ & $62\pm8$\\
DBH & 0.75 & $42.9\pm1.5$ & 0.0042 & $4.8\pm0.2$ & $5.4\pm0.8$ &
$58\pm17$\\
PMD & 0.75 & $16.9\pm0.6$ & 0.0308 & $2.8\pm0.2$ & $5.2\pm0.4$ & $58\pm13$\\
DBH & 1.5 & $26.5\pm1.5$ & 0.0121 & $3.2\pm0.1$ & $3.7\pm0.4$ & $36\pm8$\\
PMD & 1.5 & $12.2\pm1.0$  & 0.0619 & $2.1\pm0.2$ & $4.1\pm0.3$ & $45\pm7$ 
\end{tabular}
\end{ruledtabular}
\label{tab:NeTabglasses}
\end{table*}

The small drop in $N_{e}$ from melt to glass might be associated with 
changes in chain statistics.
An affine contraction of chains would not change $N_{e}$.
However, while the density increases by 15\% during the quench, the backbone
bond length $l_{0}$ decreases by only 0.5\%.
Chains in the glass are therefore stretched at short scales $(n < 10)$ 
relative to an affine contraction from the melt.
Reduction of $N_{e}$ due to this short-scale stretching would be consistent 
with our results for poorly equilibrated melts.
Additional entanglements may be created as chain ends push outward during the
quench.

Another possible explanation could be that the change in density couples to 
chain-thickness effects associated with the PPA.  
For states of constant density, increasing the bead diameter during 
implementation of the PPA decreases the calculated value of $N_{e}$.
The values of $N_{e}$ in the glassy states average $94\%$ of those in the 
corresponding melts, and $(\rho_{glass}/\rho_{melt})^{-1/3}$ is also $0.94$.
The increase in density from melt to glass could correspond to an increase in 
the effective chain thickness, which would be consistent with the observed 
reduction in $N_{e}$.
We leave further discussion of this issue to a forthcoming paper.

Values of $N_{e}$ in polymer glasses are often inferred from measurements of 
the plateau shear modulus ($G_{N}^{0} = 4\rho k_{B}T/5N_{e}$) in the 
rubbery regime just above $T_{g}$ \cite{haward97}.
Below the glass transition, rheological measurements become impossible, but
$N_{e}$ can be measured via the craze extension ratio 
$\lambda = \rho_{0}/\rho_{craze}$ \cite{kramer90}, where $\rho_{0}$ is the 
density of the undeformed glass and  $\rho_{craze}$ is the density of a 
stable craze.  The assumption that entanglements act like chemical crosslinks 
leads to the prediction 
$\lambda = \lambda^{th} \equiv \sqrt{N_{e}l_{0}/l_{K}}$, where $l_{K}$ 
is the Kuhn length in the glass \cite{kramer90}.  
Experimental and simulation values for $\lambda$ at $T \ll T_{g}$ are 
consistent with this prediction for values of $N_{e}$ obtained from the 
plateau modulus \cite{haward97,rottler03}.
However, the prediction has not been tested for direct measurements of $N_{e}$ 
in the glassy state.

Table \ref{tab:NeTabglasses} compares values of $\lambda^{th}$ predicted by
values of $N_{e}$ from PPA analysis of the undeformed glassy states to directly
measured values of $\lambda$.
The latter were obtained by straining uniaxially at constant velocity and 
measuring the ratio of the densities of coexisting uncrazed and crazed 
regions as in Ref.\ \cite{rottler03}. 
The values of $\lambda^{th}$ and $\lambda$ are consistent for the DBH states, 
but not for PMD states 
\footnote{Our values of $\lambda$ for the PMD states are slightly higher than 
those reported in Ref. \cite{rottler03} for the semiflexible-chain systems, 
perhaps because of a different technique used to measure 
$\rho_{0}/\rho_{craze}$.}.
Indeed, values of $\lambda$ are relatively insensitive to equilibration, which
suggests that the craze extension ratio is controlled mainly by large-scale 
chain structure.
Chains are greatly stretched as they pass from undeformed regions into the 
craze, and this may remove the memory of short-scale stretching from the 
fast pushoff.

\section{Discussion and Conclusions}
\label{sec:conclude}
We have shown that the density of entanglements in a model polymeric melt can
depend very strongly on equilibration of chain structure at short length 
scales. 
Values of the entanglement length for poorly equilibrated states prepared using
non-crossing chain dynamics (PMD) were found to be as much as four times
lower than values for well-equilibrated states prepared using an algorithm which
changes chain connectivity (DBH).
Coil-packing models, which focus on large-scale chain structure, fail to 
predict the magnitude of the differences in $N_{e}$.
Instead, the low values of $N_{e}$ for poorly equilibrated states 
(Table \ref{tab:NeTabflexsemiflex}) can be attributed to local chain stretching
caused by the fast pushoff procedure used to introduce excluded volume
interactions.

At the conclusion of the fast pushoff, chain retraction ensues.
Values of $N_{e}$ can increase rapidly towards apparent equilibrium over time
scales comparable to the Rouse time as short-scale chain structure equilibrates
(Table \ref{tab:equilibtimedep}, Fig.\ \ref{fig:chainstats-timedependence}).
However, large-scale chain structure remains far from equilibrium at 
$\tau_{R}$, and $N_{e}$ may continue to evolve until the chains have 
vacated their original tubes after a time $\tau_{d}$.
Since this time scales as $(N/N_{e})^{3}$ for noncrossing chains, preparation 
methods which either entirely avoid producing the local chain stretching 
(e.g., through use of chain prepacking followed by a ``slow'' pushoff 
\cite{auhl03}), or accelerate equilibration by altering chain connectivity 
\cite{karayiannis02,auhl03}, are far more suitable for the preparation of 
equilibrated states for primitive path analysis.

Changes in $N_{e}$ upon cooling from melts well above $T_{g}$ to glassy states
well below $T_{g}$ were small for all equlibration protocols and chain 
stiffnesses examined.
We did find that values of $N_{e}$ were uniformly lower in the glasses than 
in the melts; this could be attributed either to chain-thickness effects in the 
implementation of the PPA or to the stiffness of backbone bonds relative to 
intermolecular bonds.
However, the differences were comparable to the uncertainties in our 
measurements, and our results are consistent with the common assumption that
the value of $N_{e}$ in glassy states is inherited from the melt.
Finally, we found that values of $N_{e}$ inferred from measurements of the 
craze extension ratio were much less sensitive to equilibration than the 
values of $N_{e}$ obtained from PPA analyses of undeformed glassy states.
This may indicate that deformation into the craze removes any memory of the
short-scale chain stretching.

\section{Acknowledegements}
The simulations in this paper were carried out using the LAMMPS molecular 
dynamics software (http://www.cs.sandia.gov/$\sim$sjplimp/lammps.html).   
Sathish K. Sukumaran, J\"org Rottler, Kurt Kremer, and especially Gary S. 
Grest provided initial states and helpful discussions.
Support from NSF grant DMR-0454947 is gratefully acknowledged.

\end{document}